\documentclass{article}
\usepackage{arxiv}

\usepackage[utf8]{inputenc} 
\usepackage[T1]{fontenc}    
\usepackage{hyperref}       
\usepackage{url}            
\usepackage{booktabs}       
\usepackage{amsfonts,amsmath}       
\usepackage{nicefrac}       
\usepackage{microtype}      
\usepackage{graphicx}
\usepackage{natbib}
\usepackage{doi}
\usepackage{sidecap}
\usepackage{xcolor}
\usepackage{caption}
\usepackage{subcaption}

\title{12 Labours tools for developing Functional Tissue Units}

\date{June, 2024}	

\author{ \href{https://orcid.org/0000-0000-0000-0000}{\includegraphics[scale=0.06]{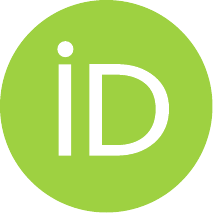}\hspace{1mm} Jagir R. Hussan}\\
	Auckland Bioengineering Institute\\
	University of Auckland\\
	Auckland, New Zealand\\
	\texttt{r.jagir@auckland.ac.nz} \\
}

\renewcommand{\headeright}{Technical Report}
\renewcommand{\undertitle}{Technical Report}
\renewcommand{\shorttitle}{FTU tools}

\hypersetup{
pdftitle={12 Labours tools for developing Functional Tissue Units},
pdfsubject={q-bio.NC, q-bio.QM},
pdfauthor={Jagir R. Hussan},
pdfkeywords={Modelling, Systems Biology, Port-Hamiltonians},
    colorlinks=true,
    linkcolor=blue,
    filecolor=magenta,      
    urlcolor=cyan,
}

\begin{document}
\maketitle

\begin{abstract}
A brief introduction of the technical approach to model FTUs as an aggregate of cells, whose state transition dynamics are mathematically represented as port-hamiltonians or Differential Algebraic equations is presented. A python library and browser based tool to enable modellers to compose the FTU graph, specify the cellular equations and the interconnection between the cells at the level of physical quantities they exchange consistent with the technical approach is discussed.
\end{abstract}


\section{Introduction}
Modelling the human physiology is a complex task. The complexity is not just limited to characterising the physiological processes in a quantifiable framework, but managing the models, connecting them and reusing them to represent the multiscale, spatio-temporal processes.

This report aims to guide modelers to acquaint with the concepts related Functional Tissue Units (FTUs), and a framework to create FTUs based on these concepts. Essential aspects related modelling and mathematical frameworks (like Port-Hamiltonians) are introduced. References are provided to support further exploration by interested readers.

\section{Functional Tissue Units}
\label{sec:headings}
\begin{SCfigure}[][h]
    \centering
    \includegraphics[width=0.4\linewidth]{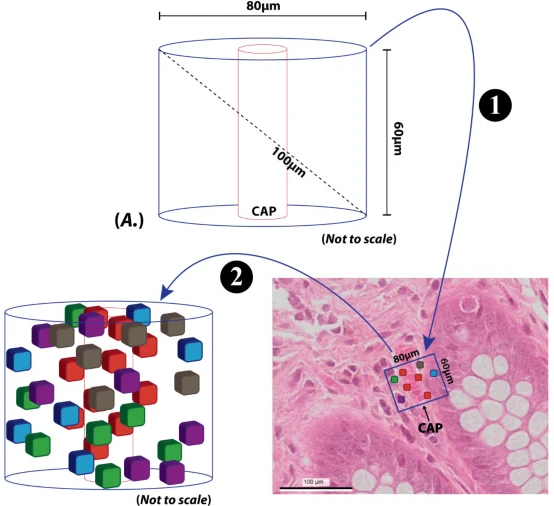}
    \caption{ The FTU template (A) according to metabolic and paracrine communication biophysical constraints-  the long axis of the resulting cylindrical block of tissue is that of the feeding capillary (CAP) on which it is metabolically dependent. This template is applied to an appropriate volumetric region in the three-dimensional histology image dataset (B). The various cells within this region (coloured boxes) are identified and their position recorded. Adpoted Figure 1 of \cite{deBono2013} }
    \label{fig:ftutemplate}
\end{SCfigure}

The notion of a Functional Tissue Unit was first introduced in \cite{deBono2013} with the aim of discretising physiological constructs, such as solid tissues in a functionally coherent manner to recreate the tissue from these units. Briefly, \emph{a FTU consists of a cylindrical diffusive field with an 80-micron diameter and 60-micron length. The long axis of this field is aligned on a central advective vessel.}

These physical dimensions are based on the observation that the group of cells are 
(i) metabolically dependent on the same capillary, 
(ii) the cellular substrate is within the limits of tissue-level molecular pathways to be coordinated via paracrine communication.

\begin{figure}
    \centering
    \includegraphics[width=0.95\linewidth]{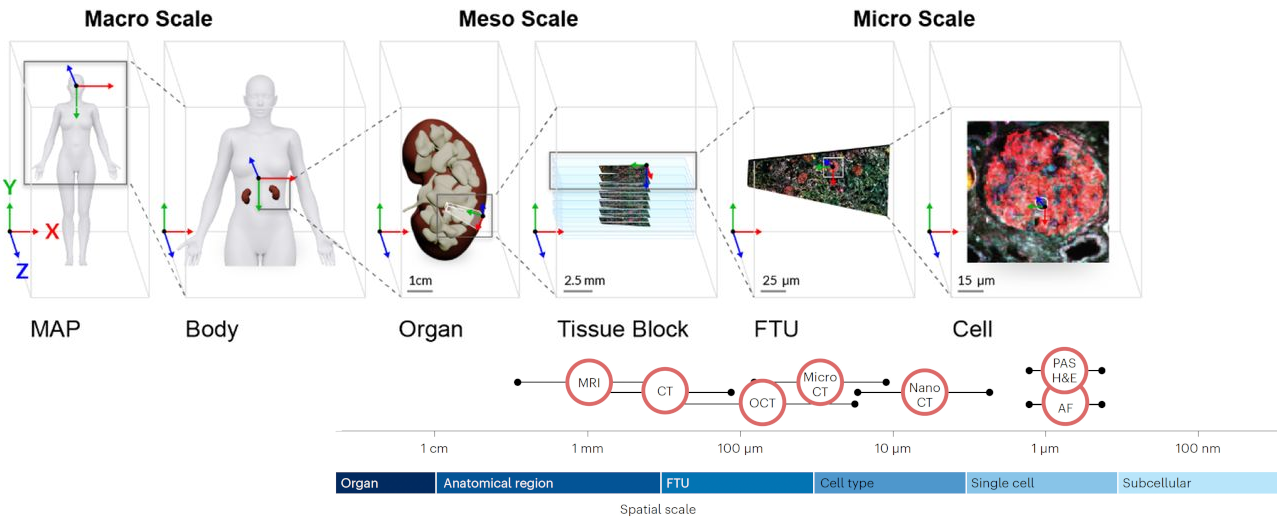}
    \caption{ Spatial scales of various physiological structures and imaging modalities through which these structures could be investigated. A example mapping of these structures for kidneys. Acronyms: CT, computed tomography; MRI, magnetic resonance imaging; OCT, optical coherence tomography; PAS, Periodic Acid-Schiff; H\&E, hematoxylin and eosin; AF, autofluorescence. Adapted from Figure 1 of \cite{Jain2023} and Figure 4 of \cite{Borner2022}.} 
    \label{fig:ftuspatialextent}
\end{figure}

In spatial terms FTU's are in the order of few hundred $\mu$m, Figure \ref{fig:ftuspatialextent}. 

The above definition provide a dimensional analysis based physiological grouping of cells. Similar principles can be used guide the identification of larger cellular organisations which will be referred to as Secondary Functional Tissue units (sFTU).
The primary organising principles that I am using to identify sFTUs, in the order of increasing spatial scale, are
\begin{enumerate}
\item Function - the physical function that the spatially local tissues is expected to provide (sFTUs could amplify the function of the primary FTU, or combine functions of different primary FTU to generate new function),
\item Structure - the agency of the group of cells to organise and remain in a connected multicellular structure,
\item Control - the ability of an agent (local and/or systemic) to control the structure and function of the multicellular organisation.
\end{enumerate}

Often there is no clear separation between these roles and requires judicious+informed approach to identify sFTUs. Typically, there are 10's to few 100's cells in an sFTU.

\section{Digital representation}
\begin{figure}[h!]
\renewcommand\thesubfigure{\arabic{subfigure}}
    \centering
    \begin{subfigure}[b]{0.49\textwidth}
         \centering
         \includegraphics[width=\textwidth]{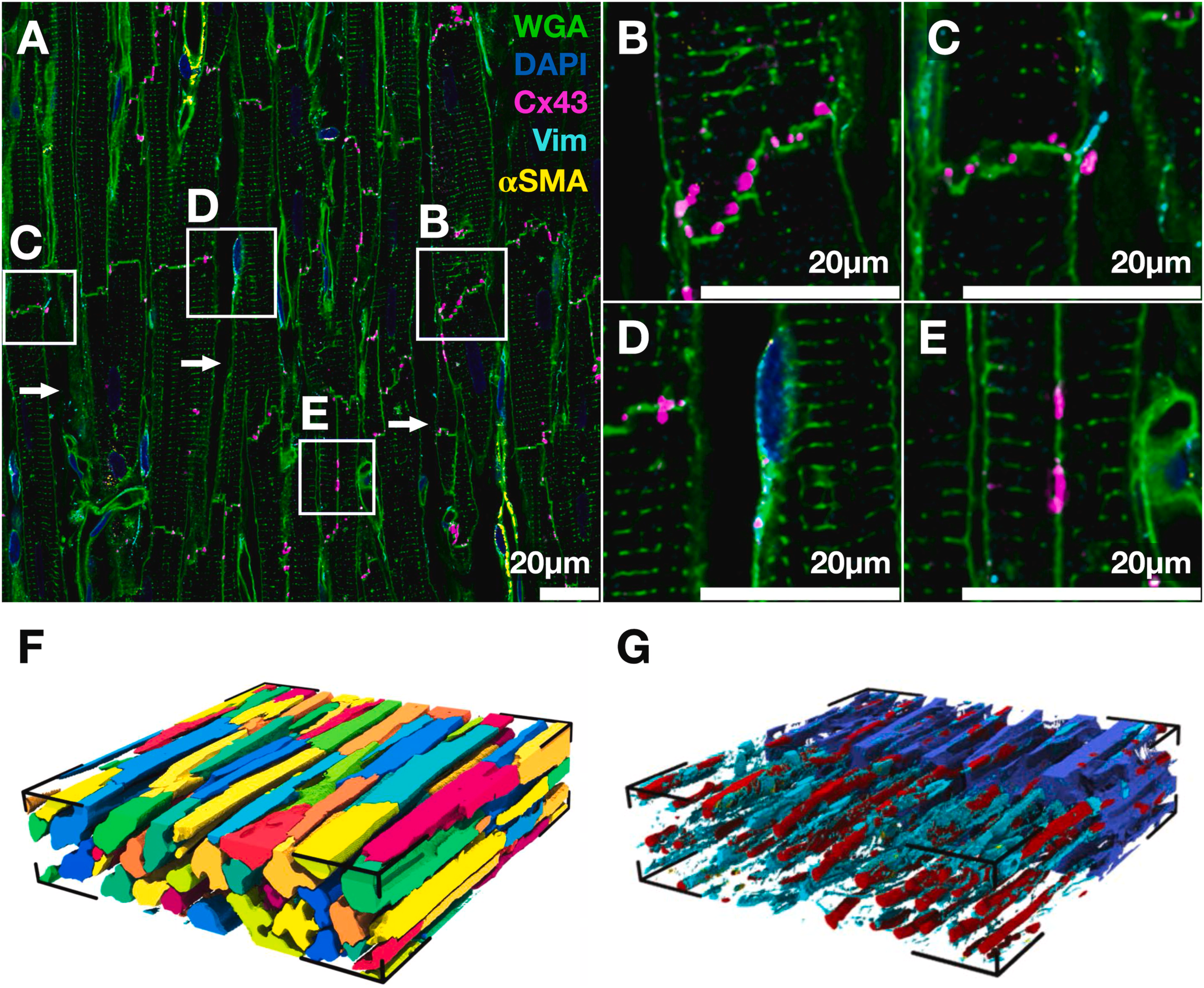}
         \caption{Normal tissue}
         \label{fig:cnorm}
     \end{subfigure}
     \hfill
     \begin{subfigure}[b]{0.49\textwidth}
         \centering
         \includegraphics[width=\textwidth]{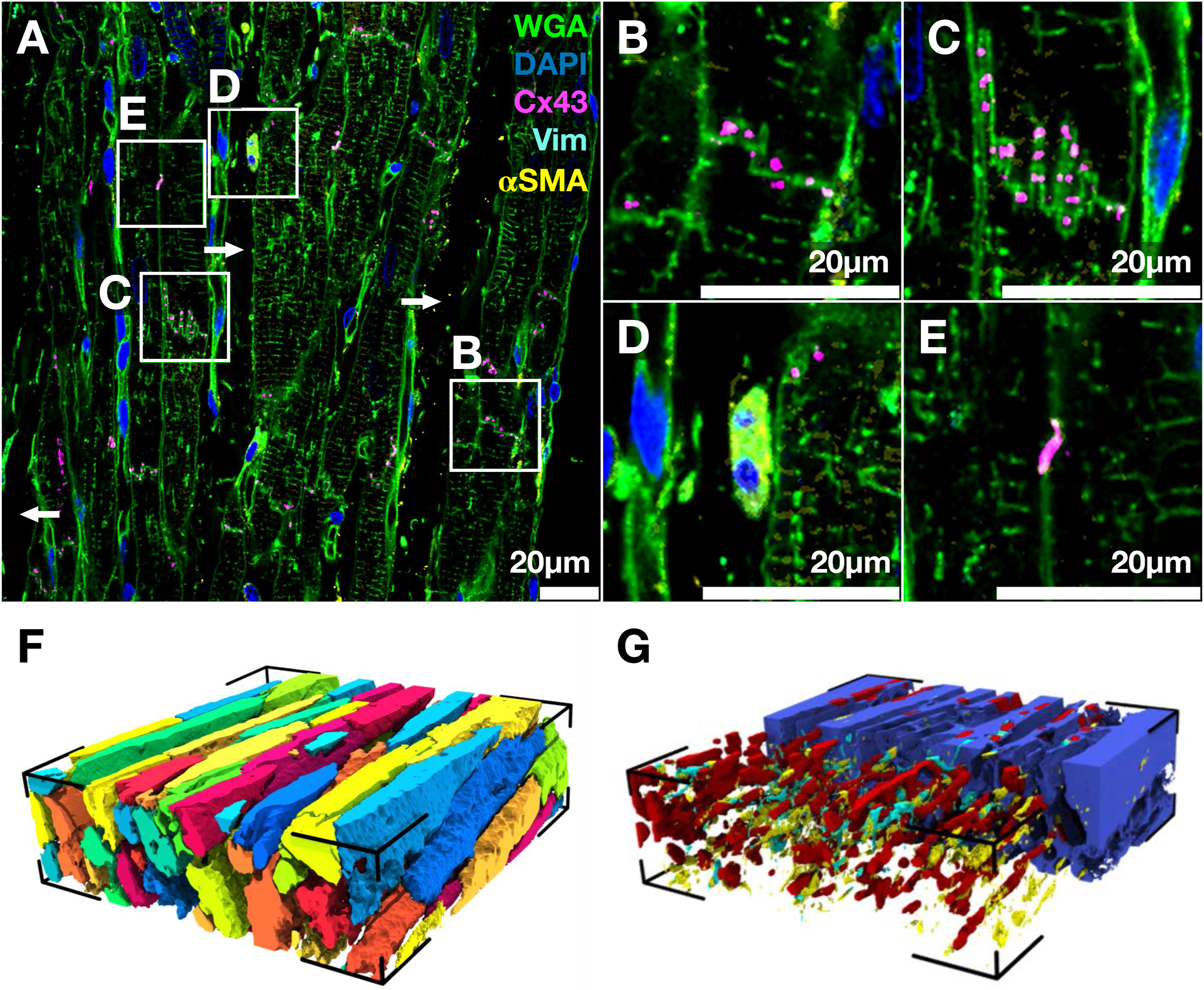}
         \caption{Infarcted tissue}
         \label{fig:cmi}
     \end{subfigure}
     \begin{subfigure}[b]{1.0\textwidth}
         \centering
         \includegraphics[width=\textwidth]{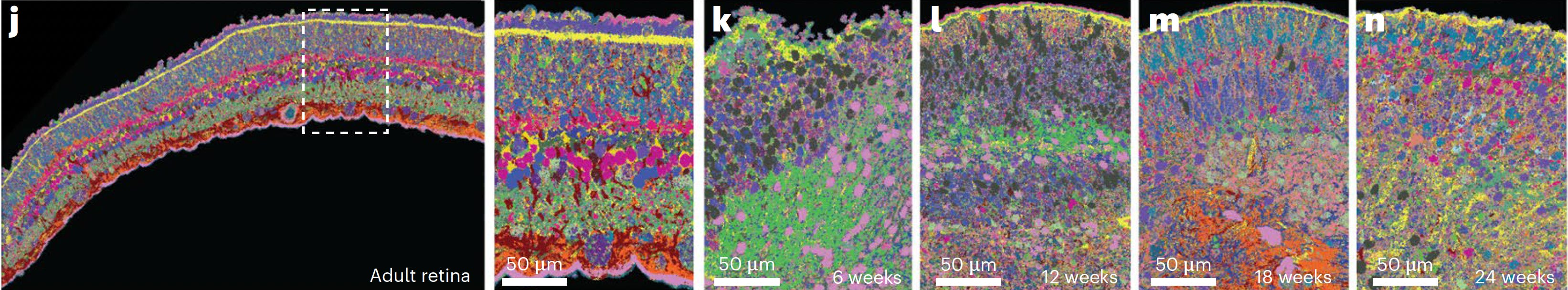}
         \caption{Retinal remodelling}
         \label{fig:cmi}
     \end{subfigure}  
     \begin{subfigure}[b]{0.9\textwidth}
         \centering
         \includegraphics[width=\textwidth]{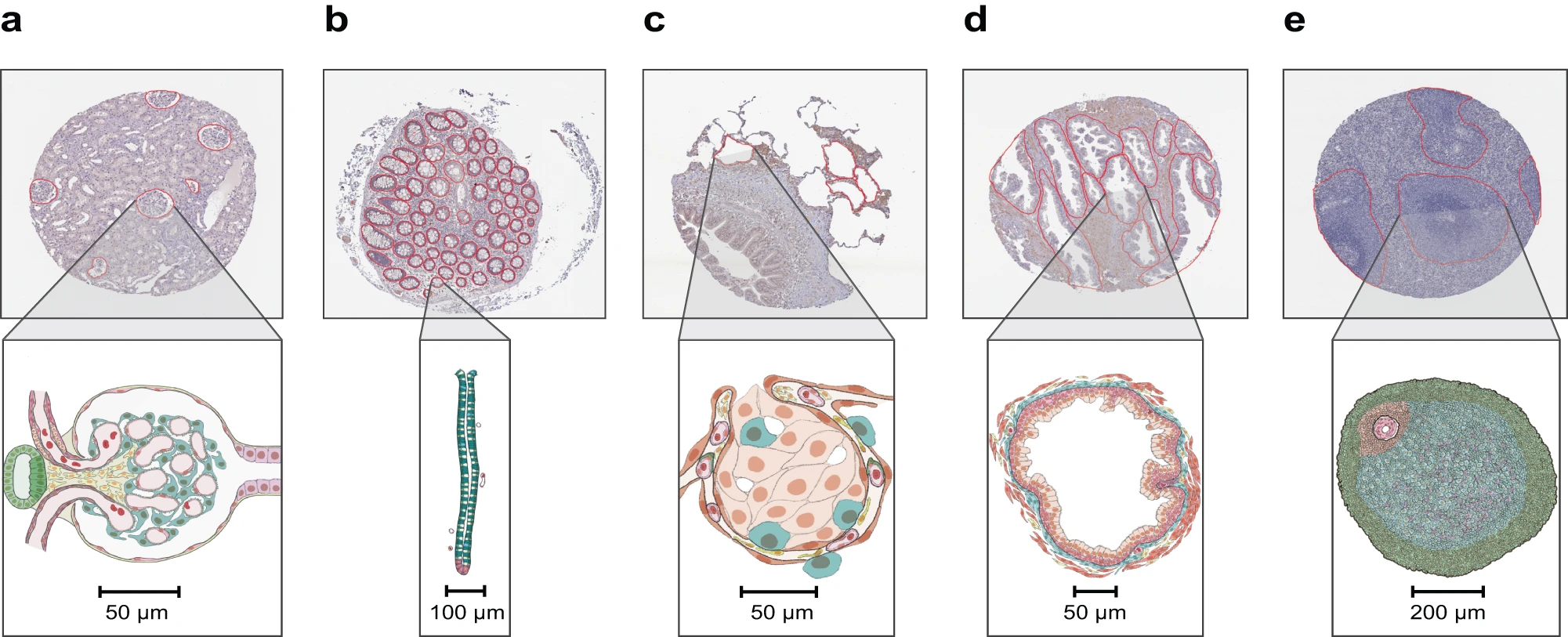}
         \caption{HubMAP segmentations}
         \label{fig:cmi}
     \end{subfigure}       
    \caption{ (1), (2): Composition and 3D reconstruction of normal and infarcted myocardium.  [B–E] Zoom-ins for the white boxes in (A) display representative microstructural features. (F) Reconstructed myocytes from the image stack. Myocytes were colored individually. The black line segments mark the edges and corners of the image stack. (G) Reconstructed non-myocyte contributors to the myocardium with ES in blue, vessels in red, fibroblasts in cyan, and myofibroblasts in yellow. The ES visualization is clipped in half for visibility of other tissue constituents.\\
    (3) Primary adult human retina section \textbf{j} and retinal organoid sections from different timepoints \textbf{k}–\textbf{n}. 
    \\
    (4) FTU segmentations by HubMAP project (a) Glomerulus in the kidney. b Crypt in the large intestine (top: perpendicular cross-section, bottom: lengthwise cross-section). c Alveolus in the lung. d Glandular acinus in the prostate. e White pulp in the spleen. 
    \\
    (1), (2) reproduced figure 1 and 2 of \cite{Greiner2022}, (3) reproduced figure 1 of \cite{Wahle2023}, (4) reproduced figure 2 of \cite{Jain2023b}}
    \label{fig:ftustaining}
\end{figure}

Once a tissue segment is identified, cellular organisation, cell types within the tissue matrix and physical mechanisms by which they communicate with each other could be determined through detailed imaging studies \ref{fig:ftustaining}  or hypothesized.

\begin{figure}
    \centering
    \includegraphics[width=0.95\linewidth]{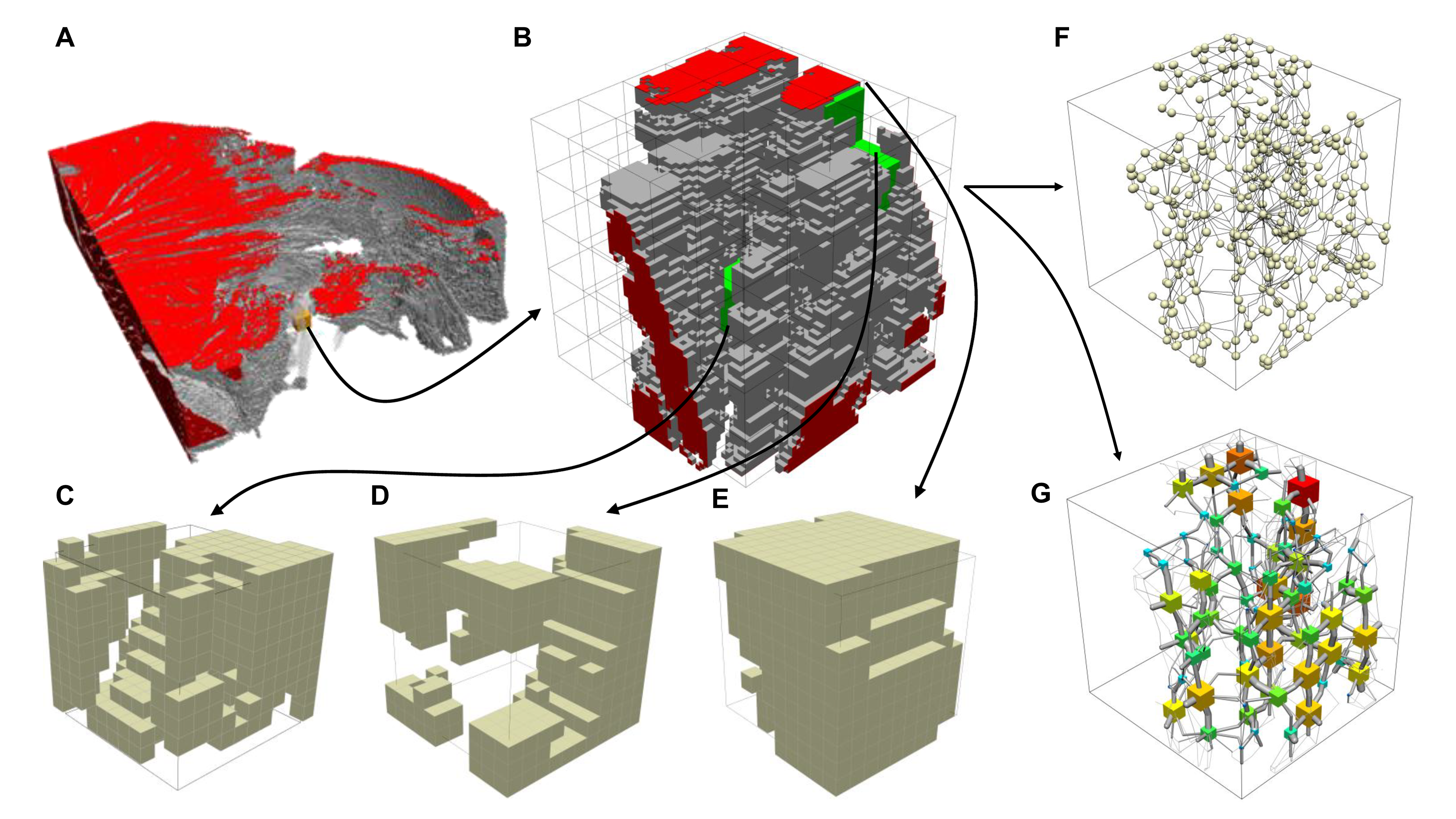}
    \caption{Schematic of the process of identifying a representative region to model a sFTU, extracting cell type information, creating cell type blocks (C,D,E) and creating a graph based representation for each block. Matrix effects can be encoded in the graph via edge annotations that characterise the influence of the matrix or via pseudo cells that capture the local influence of the matrix and integrate them with the grpah (G). Image credit. Dr Mark Trew.}
    \label{fig:ftugraph}
\end{figure}

This information along with local and non-local interaction information can then be encoded in a graph structure, Figure \ref{fig:ftugraph}. Note that matrix effects could be coupled with the graph structure by adding annotations to the edge (this could be processes like conductivity etc.) or by incorporating additional nodes(pseudo cell types) that capture local matrix effects. 

Essential characteristics that need to be captured while digitisation are:
\begin{enumerate}
    \item Identify all cell types within the FTU,
    \item Identify all the intra-cellular interactions that are active (these include electrical, mechanical, chemical, and metabolic) between these cells. Note that the list of all possible interactions between two cell types can be determined apriori and may be common across many different tissues. Therefore, determining the active subset of interactions is required,
    \item Determine the nature of the tissue matrix and the role it plays in the emerging function,
    \item All interactions (except mechanical) in physiological systems are chemical in nature; in some instances the molecules are actively transported while in most instances the molecules diffuse. Further, the timescales of the molecular diffusion also needs to be considered. 
    \item Decide for each know interaction whether is non-dissipative (follows Kirchhoff's current law for interactions), or dissipative.
\end{enumerate}
    
\section{Mathematical representation}
Computer simulation to reproduce the observed phenomena requires the digital representation to be translated to a mathematical representation that can be numerically evaluated.
Typically, this would be a system of differential (algebraic) equations describing the state transition behaviour of the cells. There are two major sources of these equations - Bondgraphs/Port-Hamiltonians and general formulations. 

Bondgraphs/Port-Hamiltonians provide a framework to represent phenomena such that physical laws (energy balance, mass/charge conservation etc.) are conserved. This enables the development of algorithmic approaches to create larger models from a set of models and ensure the resulting model also conserves physical laws.

In case of general formulations, which are unavoidable, such as phenomenological models and model reductions care must be exercised to track the influence of the non-physical operations.

The sFTU modelling framework requires a standardised representation of dynamical systems to compose sFTUs. The representation is strongly influenced by the generic cell model developed by P1 team. Briefly, the generic cell model uses bondgraphs to represent the various sub-cellular processes in a modular and physically consistent manner. It is envisioned that modellers would personalise the generic cell model to a specific cell type and use this representation in creating the sFTU.

\subsection{Dynamical systems representation}
The FTU composition framework uses the following representation for dynamical systems:
\begin{align}
    \boldsymbol{E} \dot{\boldsymbol{x}} = \left(\boldsymbol{J} - \boldsymbol{R}\right) \boldsymbol{Q} \boldsymbol{x} + \boldsymbol{B} \boldsymbol{u} \nonumber \\ 
    \boldsymbol{y} = \boldsymbol{B}^T \boldsymbol{Q} \boldsymbol{x} \label{eq:phsgeneric}
\end{align}
or equivalently
\begin{align}
    \boldsymbol{E} \dot{\boldsymbol{x}} = \boldsymbol{A} \boldsymbol{Q} \boldsymbol{x} + \boldsymbol{B} \boldsymbol{u} \nonumber \\ 
    \boldsymbol{y} = \boldsymbol{B}^T \boldsymbol{Q} \boldsymbol{x} \label{eq:pseudophsrep}
\end{align}
where $\boldsymbol{A} = \left(\boldsymbol{J} - \boldsymbol{R}\right)$.

Here $\boldsymbol{x}$ is a $n$-dimensional state vector, $\boldsymbol{J}$ is a $n\times n$ skew-symmetric matrix, $\boldsymbol{R}$ is a $n\times n$ symmetric matrix, $E$ is a $k \times n$ matrix and $Q$ is a $n\times n$ matrix, $\boldsymbol{u}$ is a $m$-dimensional input vector, $\boldsymbol{y}$ is a $m$-dimensional output vector, $\boldsymbol{B}$ is a $n \times m$ dimensional feedback matrix.

\begin{figure}
    \centering
\begin{subfigure}[b]{0.3\textwidth}
         \centering
         \includegraphics[width=\textwidth]{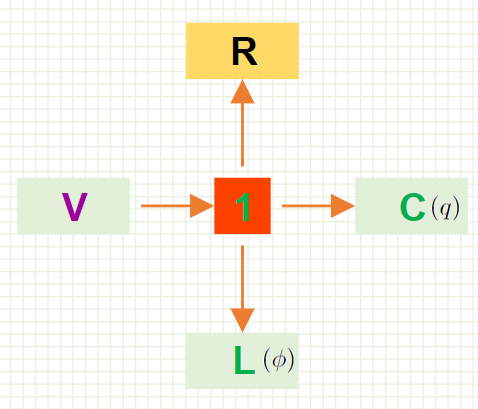}
         \caption{Bondgraph}
         \label{fig:cnorm}
     \end{subfigure}
\begin{subfigure}[b]{0.4\textwidth}
         \centering
         \includegraphics[width=\textwidth]{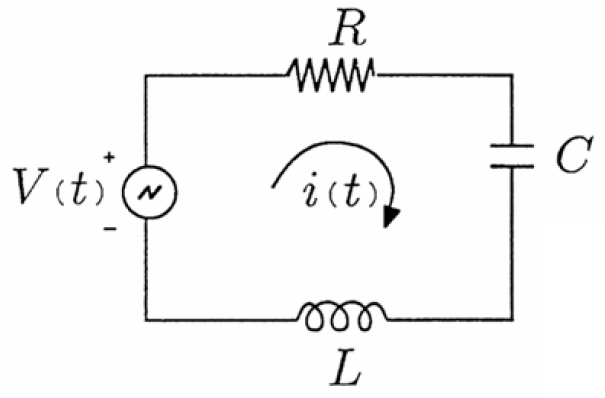}
         \caption{Circuit diagram}
         \label{fig:cnorm}
     \end{subfigure}
     \hfill
\begin{subfigure}[b]{0.9\textwidth}
         \centering

        \begin{align*}
            \begin{matrix} \frac{dq}{dt} \\ \frac{d\phi}{dt} \end{matrix} = \begin{pmatrix} 0 & \frac{\phi}{L}  \\ -\frac{q}{C} & -R\frac{\phi}{L} \end{pmatrix} + \begin{pmatrix} 0 \\ 1 \end{pmatrix} \begin{bmatrix} V\end{bmatrix}, \\
            \underbrace{\begin{bmatrix} 1 & 0 \\ 0 & 1\end{bmatrix}}_{\boldsymbol{E}} \frac{d}{dt} \underbrace{\begin{bmatrix} q \\ \phi \end{bmatrix}}_{\boldsymbol{x}} = \left( \underbrace{\begin{pmatrix}
                0 & 1 \\ -1 & 0
            \end{pmatrix}}_{\boldsymbol{J}}  - \underbrace{\begin{pmatrix}
                0 & 0 \\ 0 & R
            \end{pmatrix}}_{\boldsymbol{R}} \right) \underbrace{\begin{pmatrix}
                \frac{1}{C} & 0 \\ 0 & \frac{1}{L}
            \end{pmatrix}}_{\boldsymbol{Q}} \begin{bmatrix} q \\ \phi \end{bmatrix} + \underbrace{\begin{bmatrix} 0 \\ 1 \end{bmatrix}}_{\boldsymbol{B}} \begin{bmatrix} V\end{bmatrix}, \\
            \mathcal{H}(\boldsymbol{x}) = \frac{1}{2}\boldsymbol{x}^T\boldsymbol{E}^T\boldsymbol{Q}\boldsymbol{x}  =  \frac{1}{2}\left( \frac{q^2}{C} + \frac{\phi^2}{L} \right).
        \end{align*}
        
         \caption{State space representation}
         \label{fig:cnorm}
     \end{subfigure}     
    \caption{RLC circuit with a voltage source with potential $V$, state variables $q$-charge stored in the capacitor and $\phi$-magnetic flux in inductor. Its bondgraph, circuit diagram, state space and port-hamiltonian representation.}
    \label{fig:examplephs}
\end{figure}

Equation (\ref{eq:phs}) is a mathematical representation of a port-hamiltonian. The energy i.e the Hamiltonian, $\mathcal{H}(\boldsymbol{x})$, of the system is given by the quadratic form
\begin{align*}
    \mathcal{H}(\boldsymbol{x}) = \frac{1}{2}\boldsymbol{x}^T\boldsymbol{E}^T\boldsymbol{Q}\boldsymbol{x}
\end{align*}
Bondgraphs are a graphical representation of a subset of port-hamiltonians, as such can be directly converted to port-hamiltonian form. Tools to automatically convert bondgraphs (and the generic cell model) to port-hamiltonians are available.
A simple example demonstrating these linkages and the port-hamiltonian matrices is shown in Figure \ref{fig:examplephs}.

The port-hamiltonian matrices have physical semantics and describe energy partitioning and dissipation within the system and are useful to analysing the system's behaviour. Interested readers are referred to \cite{Schaft2014,Mehrmann2023} and references therein for an indepth analysis and discussion of port-hamiltonians.

\section{Theoretical framework to compose an FTU from port-hamiltonian like representations}
This sections provides a background to the theoretical framework to compose an FTU from cellular models - represented as port-hamiltonians, equation (\ref{eq:phsgeneric}), or with a structure that matches equation (\ref{eq:pseudophsrep}).
The following approach is adopted to ensure that the composed system continues to conserve physical laws that its constituent components conserve. 

Essentially, one is expected to connect the constituent components in such a manner that energy flows follow Kirchhoff's laws (mass/charge conservation and energy balance), and the system continues to remain passive i.e. it does not produce energy, it can only dissipate or conserve. Interested readers are referred to \cite{Jeltsema2003} for technical details.

\begin{figure}
    \centering
    \includegraphics[width=0.5\linewidth]{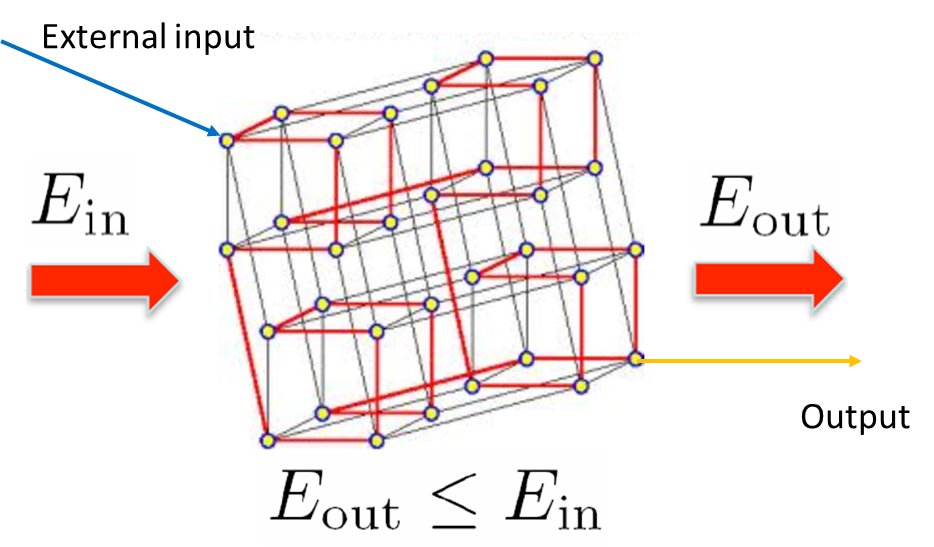}
    \caption{Schematic of energy flow through interconnected systems (nodes) and the concept of passivity that requires the energy that flows out of the system is less than or equal to the energy that flows into the system. The composite system receives external inputs from a subset of nodes and outputs/communicates with the environment through anothet subset of nodes (not mutually exclusive from the ones from which it recieves input).}
    \label{fig:phspassivity}
\end{figure}
\subsection{Notation: Port-Hamiltonian Differential Algebraic Equation (PH-DAE)}
We setup some notation to support the discussion in the following section.\\
A $n$-dimensional differential-algebraic equation of the form
\begin{eqnarray}
    \frac{d\,Ex}{dt}&=&(J-R)Qx + Bu, \label{eq:phsnot} \\
    y&=&B^TQx. \nonumber
\end{eqnarray}
is called a port-hamiltonian differential-algebraic equation (PH-DAE), if the following are satisfied:
\begin{enumerate}
\item There exist linear transformations of the form
\begin{equation*}
    E \in \mathbb{R}^{k\times n},\quad J, R, Q \in \mathbb{R}^{n\times n}, \quad B \in \mathbb{R}^{n\times m}. 
\end{equation*}

\item There exists a subspace $\mathcal{V}\subset\mathbb{R}^n$ with the following properties:

\begin{enumerate}
\item for all intervals $\mathcal{I}\subset\mathbb{R}$ and functions $u:\mathcal{V}\rightarrow\mathbb{R}^m$ such that eq. (\ref{eq:phsnot}) has a solution $x:\mathcal{V}\rightarrow\mathbb{R}^n$, it holds $Qx\in \mathcal{V}$ for all $t\in \mathcal{I}$.

\item $J$ is skew-symmetric on $\mathcal{V}$. 
\item $R$ is symmetric on $\mathcal{V}$, such that $0 \leq (Qx)^T R Qx$ for all $x\in \mathcal{V}$. 

\end{enumerate}

\item There exists some function $H\in C^1(\mathbb{R}^n,\mathbb{R})$, the Hamiltonian, such that $\nabla H(x) = E^T J Qx$ for all $Qx \in J^{-1} \mathcal{V}$.
\end{enumerate}

Port-Hamiltonian systems fulfill energy balance and the total energy of a PH-DAE at time $t$ is given by $H(Qx(t))$. The power inflow is given by the inner product of input $u$ and output $y$.
\subsection{Composing a PH-DAE from sub-systems which are PH-DAE's}
\begin{figure}
	\centering
	\includegraphics[width=0.95\linewidth]{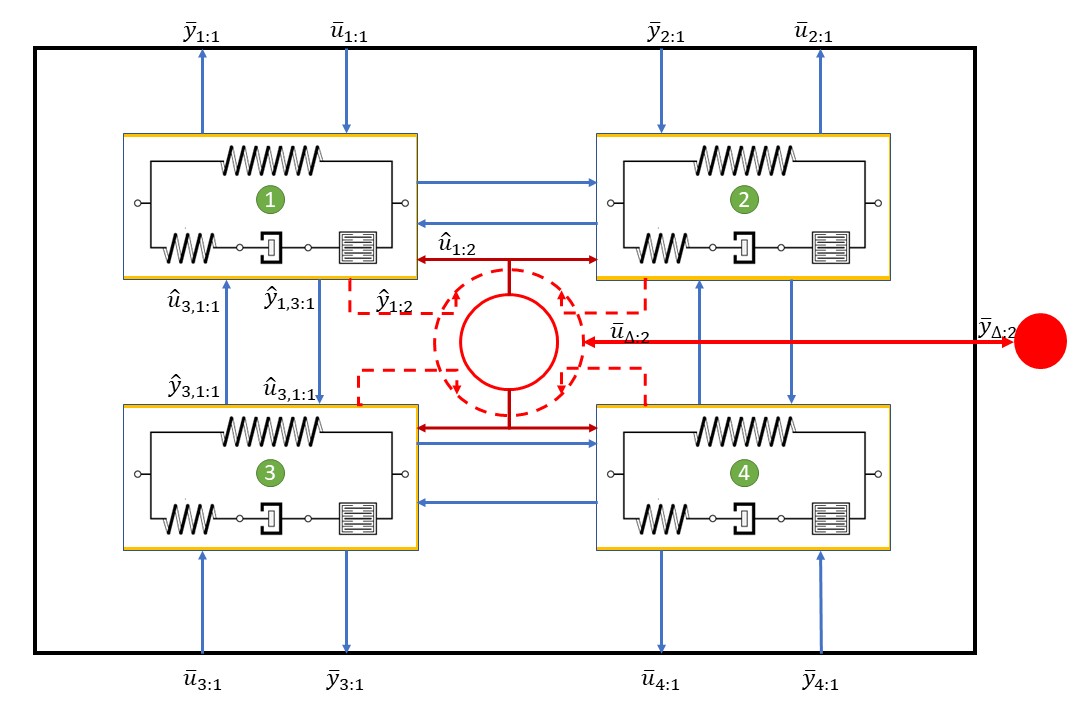}
	\caption{Schematic of a composite PH-DAE model. The composite model is made of four PH-DAE models representing a subsystem. Sub systems are interconnected with other systems (the input and output ports are indicated with a $\hat{h}$), some may also be connected with the environment (the input and output ports are indicated with a $\bar{b}$). Interconnections must match the physical units of the component that use for connecting. Sub systems can connect directly to other systems (\textcolor{blue}{blue arrows}), sub systems can also connect in a dissipative manner, where the interconnecting medium modulates the dissipation (\textcolor{red}{red arrows}). \\
 While other type of interconnections are possible, currently the above two are supported. Also, the sub systems depicted are of the same type for simplicity of presentation. Ideally, the sub systems will be of different types but share components of the same physical units. \\ 
 The input and output component labels encode the interconnection information. For instance $\hat{y}_{a,b:c}$ implies, system $a$'s output of component $c$, connects with subsystem $b$. Similarly, $\hat{u}_{a,b:c}$ implies that system $a$ receives an input for component $c$ from subsystem $b$. $\bar{y(u)}_{a:c}$ implies, system $a$ outputs to (receives input from) the environment ($\Delta$) through component $c$.}
	\label{fig:compositePHS}
\end{figure}

Synthesizing FTU's from the PH-DAE's of subsystems requires the coupling of several PH-DAE's involving different inputs and outputs, Figure \ref{fig:compositePHS} - some of which will exclusively connect to other sub-systems (nodes linked by black lines in Figure \ref{fig:phspassivity}, henceforth referred to as internal) and some of which will connect to the environment (nodes with blue and yellow arrows incident/outgoing in Figure \ref{fig:phspassivity}, henceforth referred to as external). 

The resulting interconnection topology should ensure that the physical laws are not violated and the resulting system is passive. 

The following approach to compose the compact PH-DAE from multiple PH-DAE's is based on the method outlined in \cite{gunther2021}.\\
Consider $k\geq2$ copies of PH-DAE' of the form (\ref{eq:phs}), where the indices ($i$) identify the PH-DAE.
\begin{eqnarray}
    \frac{d\, E_ix_i}{dt}&=&(J_i-R_i)Q_ix + B_iu, \label{eq:phs} \\ 
    y_i&=&B_i^TQ_ix_i. \nonumber
\end{eqnarray}
Each PH-DAE is associated with Hamiltonian $H_i$. To account for input-output coupling that are exclusively among the PH-DAE instances (internal) and input-output coupling that connect to the environment (external); the input and output vectors of each PH-DAE are split as
\begin{equation*}
    u_i(t) = \begin{bmatrix} \hat{u}_i(t) \\ \bar{u}_i(t) \end{bmatrix}, \quad     y_i(t) = \begin{bmatrix} \hat{y}_i(t) \\ \bar{y}_i(t) \end{bmatrix}.    
\end{equation*}
Here bar-accents refer to external inputs/outputs, and the hat-accents refer to internal input/output data. Correspondingly, the port matrix, $B_i$, is split as $B_i=\left[ \hat{B}_i \bar{B}_i \right]$. Note that some PH-DAE instances may have zero internal/external input/output connections.

Each input variable, $u_c \in u$ denoted $\hat{u_i}:c$, of the PH-DAE can be coupled directly or dissipatively. For directly coupled input variables, the subsystems are coupled via skew-symmetric topological coupling matrix that links the inputs ($u_c$) and outputs ($y_c$) such that the following relationship is satisfied
\begin{equation*}
    \hat{u}_i:c = \Sigma^k_{j=1,k\neq i}\: \hat{C}_{i,j} \hat{y}_j:c,\quad \forall i=1\dots k.
\end{equation*}
Here 
$\hat{C}_{i,j} \in \left\{ 1,0,-1 \right\}^{dm_i \times dm_j}$, $dm_i$ is the number of directly coupled input/output variables of the PH-DAE $i$.

For dissipatively coupled input variables, denoted $\hat{u_i}:\gamma$, the subsystems are coupled using the weighted graph Laplacian matrices, $_\gamma \hat{L}$, that link the corresponding input and output variables as
\begin{equation*}
    \hat{u}_i:\gamma = \Sigma^k_{j=1,k\neq i}\: _\gamma\hat{L}_{i,j} \hat{y}_j:\gamma,\quad \forall i=1\dots k.
\end{equation*}

The number of Laplacian matrices is determined by the number of dissipatively coupled input variables, each characterized by its own dissipation coefficients (as determined by the tissue matrix). 
Given these matrices the composite PH-DAE for the whole system can be compactly represented as
\begin{eqnarray}
    \frac{d}{dt}\begin{bmatrix}
        E& 0& 0\\
        0& 0& 0\\
        0& 0& 0
    \end{bmatrix}\begin{bmatrix}x\\\hat{u}\\\hat{y}\end{bmatrix}& =& \left ( \begin{bmatrix}
        J& \hat{B}& 0 \\
        -\hat{B}^T& 0& I \\
        0& -I& -\hat{C}
    \end{bmatrix}
    - 
    \begin{bmatrix}
        R& 0& 0 \\
        0& 0& 0 \\
        0& 0& \hat{L}
    \end{bmatrix}
    \right )\begin{bmatrix}Qx\\\hat{u}\\\hat{y}\end{bmatrix}
    +
    \begin{bmatrix}\bar{B}\\0\\0\end{bmatrix}\bar{u}, \label{eq:compositephs} \\ 
    \bar{y} &=& \begin{bmatrix}
        \bar{B}^T & 0& 0
    \end{bmatrix} \begin{bmatrix}Qx\\\hat{u}\\\hat{y}\end{bmatrix} .\nonumber
\end{eqnarray}

Here, the vectors are constructed as 
\begin{align*}
x^T = (x^T_1, \dots ,x^T_k), \\
\bar{u}^T = (\bar{u}^T_1, \dots ,\bar{u}^T_k),\\
\hat{u}^T = (\hat{u}^T_1, \dots ,\hat{u}^T_k),\\
\bar{y}^T = (\bar{y}^T_1, \dots ,\bar{y}^T_k),\\
\hat{y}^T = (\hat{y}^T_1, \dots ,\hat{y}^T_k).
\end{align*}
The matrices are constructed as 
\begin{align*}
E = \text{diag}(E_i,\dots,E_k), \\
J = \text{diag}(J_i,\dots,J_k),\\
R = \text{diag}(R_i,\dots,R_k),\\
Q = \text{diag}(Q_i,\dots,Q_k),\\
\bar{B} = \text{diag}(\bar{B}_i,\dots,\bar{B}_k),\\
\hat{B} = \text{diag}(\hat{B}_i,\dots,\hat{B}_k).
\end{align*}

The Hamiltonian of the composite system with $k$ PH-DAE instances is given by
\begin{equation}
 H = H_i+\dots+H_k. \label{eq:khamiltonians}
\end{equation}

Equation (\ref{eq:compositephs}) and (\ref{eq:khamiltonians}) provide the mathematical representation of the FTU.

\subsection{Numerical solutions}
Numerical solutions of (\ref{eq:compositephs}), involve dynamic iteration schemes (see \cite{strang22}). These schemes differ in how the internal inputs are computed. 

A \emph{Jacobi-type} approach solves the system of equations using the state values from the previous time step to determine the inputs. Here we follow such an approach and rewrite (\ref{eq:compositephs}) by explicitly indicating the time steps from which the component values are used.

\begin{eqnarray}
    \frac{d}{dt}\begin{bmatrix}
        E& 0& 0\\
        0& 0& 0\\
        0& 0& 0
    \end{bmatrix}\begin{bmatrix}x^{t+1}\\\hat{u}^{t+1}\\\hat{y}^{t+1}\end{bmatrix}& =& \left ( \begin{bmatrix}
        J& \hat{B}& 0 \\
        -\hat{B}^T& 0& I \\
        0& -I& 0
    \end{bmatrix}
    - 
    \begin{bmatrix}
        R& 0& 0 \\
        0& 0& 0 \\
        0& 0& 0
    \end{bmatrix}
    \right )\begin{bmatrix}Qx^{t+1}\\\hat{u}^{t+1}\\\hat{y}^{t+1}\end{bmatrix}
    +
    \begin{bmatrix}\bar{B}&0\\0&0\\0& -(\hat{C}+\hat{L})\end{bmatrix} \begin{bmatrix}
        \bar{u}^{t+1}\\
        \hat{y}^{t}
    \end{bmatrix}, \label{eq:phsjacobi} \\ 
    \bar{y}^{t+1} &=& \begin{bmatrix}
        \bar{B}^T & 0& 0\\
        0&0& -(\hat{C}+\hat{L})^T
    \end{bmatrix} \begin{bmatrix}Qx^{t+1}\\\hat{u}^{t+1}\\\hat{y}^{t+1}\end{bmatrix}. \nonumber
\end{eqnarray}

Consider the ode part of equation (\ref{eq:phsjacobi}) rewritten in terms of the non-zero components.
\begin{eqnarray}
    \frac{d}{dt}Ex^{t+1} &=& (J - R)Qx^{t+1} - \hat{B}(\hat{C}+\hat{L})\hat{B}^T Qx^{t} + \bar{B}\bar{u}^{t+1},   \label{eq:phsjacobire1}
\end{eqnarray}
here the following relationships are used to compute the $\hat{B} \hat{u}^{t+1}$ term:
\begin{eqnarray*}
    \hat{u}^{t+1} &=& (\hat{C}+\hat{L})\hat{y}^t, \nonumber \\
    \hat{B} \hat{u}^{t+1} &=& \hat{B}(\hat{C}+\hat{L})\hat{y}^t,  \nonumber \\
\end{eqnarray*}
since 
\begin{eqnarray*}
    \begin{bmatrix}
        \hat{y}^{t} \\
        \bar{y}^{t} \\
    \end{bmatrix} &=& [\hat{B} \bar{B}] Q x^{t}, \nonumber \\
    \hat{B} \hat{u}^{t+1} &=& \hat{B}(\hat{C}+\hat{L})\hat{y}^t,  \nonumber \\
    &=& \hat{B}(\hat{C}+\hat{L})\hat{B}^T Q \hat{x}^t. \nonumber
\end{eqnarray*}

Letting $K := \hat{B}(\hat{C}+\hat{L})\hat{B}^T$, system  (\ref{eq:phsjacobire1}) can be written as:
\begin{eqnarray}
    \frac{d}{dt}Ex^{t+1} &=& (J - R)Qx^{t+1} - K Qx^{t} + \bar{B}\bar{u}^{t+1},   \label{eq:phsjacobire}
\end{eqnarray}

and is used to iteratively solve for the state transition dynamics.

\subsection{Analytical techniques}
The symbolically resolved port-hamiltonian matrices provide opportunities to analyse the dynamics of the system and inform model development. Approaches such as dynamic mode decomposition may be used to couple the insights from simulated data to augment symbolic analysis. 

Tools and workflows to support numerical and symbolic analysis are under development and will be made available to the 12L community.
\section{FTUUtils:A python library to compose FTUs from PH-DAEs}
A python based library from composing FTUs from PH-DAE descriptions has been developed and available as open-source. The library enables modelers to export compositions into symbolic form for downstream symbolic analysis and as python code for numerical analysis. A model development pipeline based on the library is shipped with the library code.
A GUI interface to this library has also been developed and discussed in detail below.

\subsection{FTUWeaver: A browser based tool to compose FTUs from PH-DAEs}
This section describes \emph{FTUWeaver} a browser based tool to compose FTUs from PH-DAE descriptions. The tools provides an UI to create a representative graph of the FTU, load and/or create PH-DAEs in the required format, define connection networks and their properties. The tool captures this information and bundles it in a format suitable for the subsequent analysis.
The focus of the ensuing discussion is to present the features of this tool and how the concepts discussed above are digitally represented.

A instance of the tool is publicly available online at \url{http://130.216.217.109/} 

This address may change in the future.

The tool is self contained and runs on the client and does not require sustained internet connectivity once the dependent modules are loaded on to the browser. The tools is supported on Google Chrome and Microsoft Edge browsers, although it should be able to load and run on other browsers (that supports javascript and webassembly), this has not be tested and will not be supported.

\subsection{The graphical composer interface}
\begin{figure}
    \centering
    \includegraphics[width=1\linewidth]{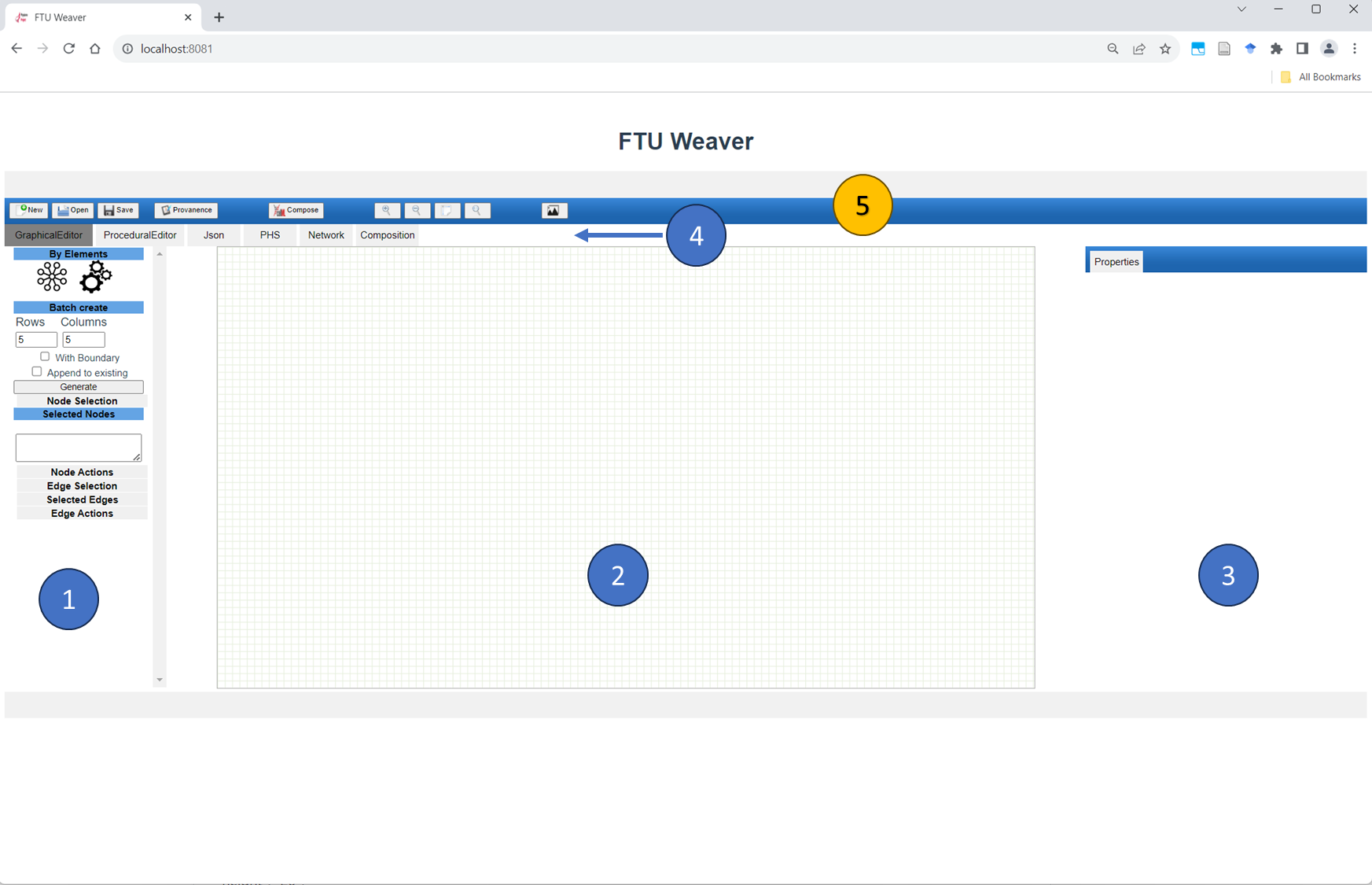}
    \caption{The opening page of the FTU Weaver tool. The dashboard consists of five major parts. 1) Toolbar, 2) Canvas, 3) Properties bar, 4) Editors and 5) Control bar. See text for more details.}
    \label{fig:ftuweaverdashboard}
\end{figure}
The graphical editor provides, Figure \ref{fig:ftuweaverdashboard}, an interface to graphically compose an FTU. The process begins with creating the backbone graph which represents the cellular components (as nodes) and their interactions/connectivity (as edges).

\subsubsection{Design concepts}
There are a few concepts required to understand the approach using which the tool encodes composite FTU information- Networks and  Boundary nodes. We discuss these in detail below.

\textbf{Networks}

Each cell type interacts with the extracellular environment, other cells and cellular matrix through one or more chemical species, electrical signals and mechanical impulses. The digital representation of the (s)FTU therefore should provide a framework for the exchange of these masses/signals and impulses.

\begin{figure}[h!]
    \centering
    \includegraphics[width=1\linewidth]{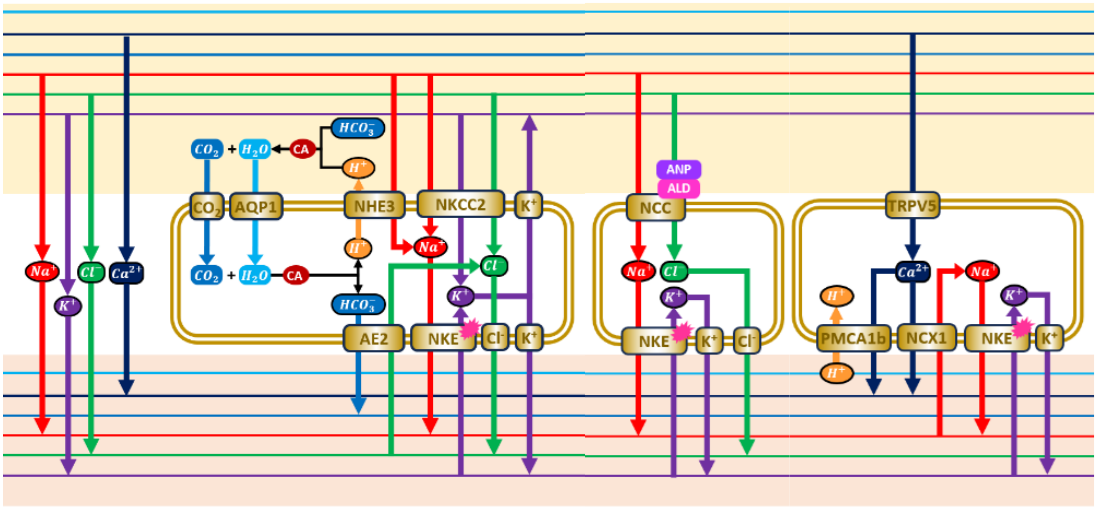}
    \caption{Model of reabsorption of electrolytes in the thick ascending limb and early distal tubule in the kidney. Networks for each chemical species are coloured differently, species could diffuse along the network or actively transported by a cell type.\\
    Image Credit: Peter Hunter}
    \label{fig:networkexample}
\end{figure}

This is implemented using the concept of networks within the digital representation of the FTU. Each network is exclusively associated with a physical quantity and allows the exchange of this physical quantity among cells that are connected to this network, Figure \ref{fig:networkexample}.  

When a graph is defined, networks are defined on the edges. Networks are identified by a number and a weight. The weight is used by dissipative networks to determine the diffusion strength. By convention all internal networks are positively numbered, boundary networks are negatively numbered. 

Networks are further classified as dissipative, non-dissipative and boundary networks. 

\emph{Dissipative networks} allow the associated quantity to be diffusively transported with a loss. The dissipation strength is specified on the edge that connects nodes. 

\emph{Non-Dissipative networks} ensure mass/charge/quantity conservation (Kirchhoff's current law). There are no strengths or edge weights associated with these networks.

\emph{Boundary networks} are a special type. They are designed to connect the FTU to external sources. They can be dissipative or non-dissipative as well. However, the amplitude of the source is constant with respect to each connected node - as if each connected internal node views a single instance of the boundary source.
Boundary networks also provide an option to specify thresholds/resting-potential etc.

\textbf{Technical implementation note}

Graph theoretic implementation of Dissipative and Non-dissipative networks to calculate $\hat{C}$ and $\hat{L}$ in equations (\ref{eq:phsjacobire1},\ref{eq:compositephs}) is as follows:

For dissipative networks, the \textit{weighted graph Laplacian} is calculated for the network and the Laplacian weights for the nodes are set for the component (input-vector element) in $\hat{L}$ matrix. 

For non-dissipative networks, the \textit{Adjacency matrix} is calculated for the network and used to populate the $\hat{C}$ matrix for that component (input-vector element). Note that $\hat{C}$ is the connectivity matrix.

\textbf{Boundary Node}

FTUs can receive inputs that are phenomenological/user-defined. Such inputs may not be mathematically represented and require an explicit data integration approach. Boundary Nodes support this approach, each node defines a boundary network that feeds this information.

\textbf{Default options}

When a lattice-graph is created a default network is created and the internal network is assigned number $1$.  When boundary is created, a network for each boundary node is created and assigned a negative number ($-\text{boundarynode.id}$)

\subsubsection{Graph generation}
This graph can be created by hand by dragging and dropping nodes on to the canvas. The UI provides two types of generic nodes, internal \includegraphics[height=15pt]{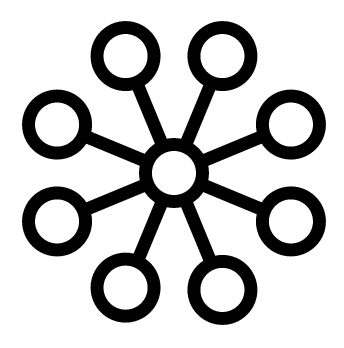} and boundary 
    \includegraphics[height=15pt]{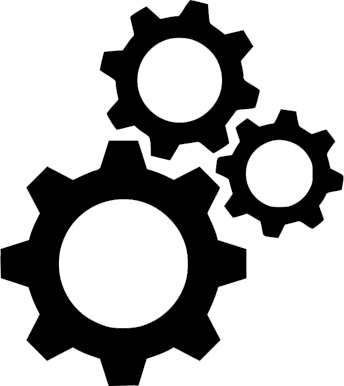}. The semantic is to highlight the behaviour of these nodes. The need for the separation and the use of boundary nodes will become clear as we proceed. 

Each node is provided an unique identifier based on the order in which they are created. The node identifier is used to programmatically select nodes.

Nodes can be connected to each other by clicking on a node and dragging the mouse cursor with the left mouse button pressed to the target node. The target node will be highlighted when a connection can be made. Releasing the mouse button will make the connection. 

The tool also provides a feature to generate 2D lattice graphs (suited for modelling syncytial tissues). Users are required to provide the number of rows and columns in the lattice. All nodes that are created are of internal type and are arranged in a grid. Users can also choose to create boundary for the lattice. In such instances, the outer nodes (of type internal) of the lattice are connected to boundary nodes. 

By default, creating a lattice using this feature will remove any current graph in the canvas and create a new lattice as requested. However, there may be instances when more than one lattice is required to model group behaviour (one lattice for excitable cells, another for passive cells with some connectivity between these lattices). The UI provides a checkbox to allow for generating multiple lattices in a sequence by appending new lattices to existing ones in the canvas.

\subsubsection{Graph manipulation}
It is likely that some nodes or edges require to be deleted or their properties need to be updated. Nodes and edges can be selected by clicking on them. The canvas also provides a rubberband feature (click on an empty part of the canvas and drag the mouse with the left mouse button pressed. This action will create a rectangular sprite that indicates the region within which all nodes and edges will be selected. Releasing the mouse button will finalise the selection).

Node and edge action controls are organised within an accordion html element. Users are expected to click on the control to access the features or to hide them.

Selected node ids and edge ids are listed inside the \colorbox{blue!30}{ Selected Nodes} and \colorbox{blue!30}{ Selected Edges} textboxes in the toolbar. Users can edit the list of nodes and edges to modify the selected items on the canvas.

The control \colorbox{blue!30}{Node actions} lists the possible node manipulations that can be performed. These include, node deletion, setting the Port-Hamiltonian description that should be used for representing its dynamics, connecting the selected nodes to a target node.

Similarly, The control \colorbox{blue!30}{Edge actions} lists the possible node manipulations that can be performed. These include, deleting edges, showing or hiding their labels and setting weights for the networks associated with the edge.

\subsubsection{Property manipulation}
Nodes and Edges have properties. Nodes essentially require a port-hamiltonian to represent the states and the transition dynamics active on the node. This property is shown in the \colorbox{blue!30}{Properties bar} when a node is selected.

Edges have one or more networks associated with them. For dissipative networks, edges are assigned a weight that is used to compute the graph Laplacian. For non-dissipative networks, the weight is disregarded.

Node and Edge properties need to be set through the \colorbox{blue!30}{Node actions} and \colorbox{blue!30}{Edge actions} controls.

\subsubsection{Canvas view manipulations}

The graph canvas can be zoomed-in,-out, fit-to-view, reset to original view - using the controls available,\includegraphics[height=15pt]{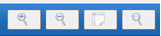} , on the Control bar.

Users can also insert images to help guide the FTU graph creation process using the background image control \includegraphics[width=15pt]{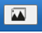}.

\subsubsection{Editors}
The User interface provides several editors to input information necessary to setup an FTU, interrogate files and also a python shell to programmatically create the graphs and set node/edge properties. These Editors are organised as tabs and can be accessed by clicking on their name in the Editors tab bar, 4 Figure \ref{fig:ftuweaverdashboard}. 
Each of these editors and their use will be introduced below in the context where the information requirement arises.

\subsubsection{Loading Port-Hamiltonians}
Central to the creation of FTUs is the ability to specify port-hamiltonians descriptions for each cell type with in the FTU. Ideally, these port-hamiltonian description will be made available by tools that allow the use to generate specific cell type from the generic cell model. However, there will instances where phenomenological models, approximations etc. need to be used. In such cases, the PHS Editor enables users to load previously created port-hamiltonian descriptions or create new descriptions.
\begin{figure}
    \centering
    \includegraphics[width=1\linewidth]{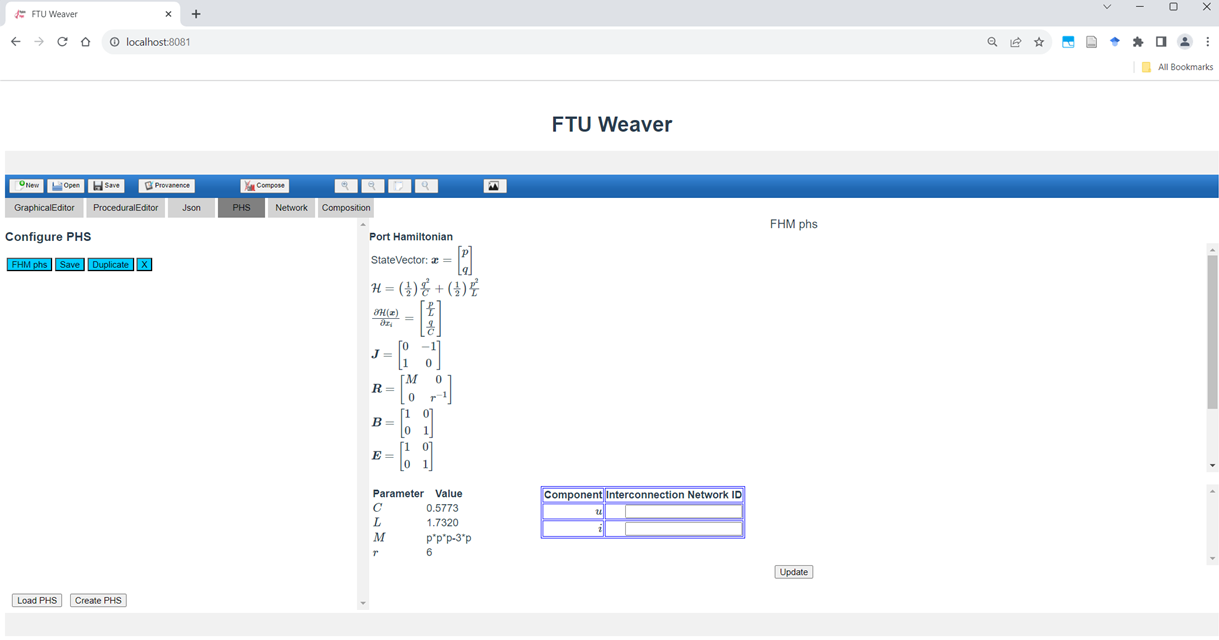}
    \caption{Port-Hamiltonian loading/creation/manipulation UI.}
    \label{fig:phseditorui}
\end{figure}

Existing port-hamiltonians can be loaded using the \colorbox{gray!10}{Load PHS} control. On clicking this control, a popup will appear to enable the selection of the phs description from disk. The user can also select a colour to be associated with the PHS entry that will be loaded. The colour information is useful for visualising the composed FTU graph, as nodes that have be assigned to use this PHS entry are assigned the colour.
The left hand side tool bar of the editor shows all the loaded/created port-hamiltonian descriptions. The control buttons associated with the instances are assigned the colour assigned to the entry.

The panel on the right hand side of the editor shows all details associated with the port-hamiltonian. Except for the parameter values and the component interconnection network information the other entries are read-only.

For each PHS entry, users are expected to provide information regarding the networks through which the inputs of the entry interact. For instance, if the input to the PHS consists of $\begin{bmatrix}
    u\\i
\end{bmatrix}$, and the input-vector component $u$ interacts on network with id $1$, the specify $1$ in the text box next to the component on the 'Interconnection Network ID' column. Similarly specify the networks for other components of the input vector.

It is a good practice to click on the \colorbox{gray!10}{Update} control when changes are made to the parameters or network assignments.

\textbf{Parameter variation}

It is very likely that some cell types share similar state transition logic, but differ in the parameters that guide the logic. In such cases, there is a single PHS description that is shared by these cell types, except for the parameters associated with the cell type. This is supported by the \colorbox{cyan!10}{Duplicate} control associated with a PHS definition. On clicking this control, a popup will appear to specify a name for the duplicate and select a colour associated with the new entry. The new entry is then created and listed on the left hand side.

\colorbox{cyan!10}{Save} control will save the phs entry to disk (to the downloads directory). The \colorbox{cyan!10}{X} control will remove the phs entry from the list of available phs.

\textbf{Creating new PHS definitions}

New PHS definitions can be created using the \colorbox{gray!10}{Create PHS} control. On clicking this control a popup will appear to obtain the PHS information and convert it into a format that the UI tool understands. Currently, placeholder data is provided to help the user with the input syntax. All vectors and matrices are input using the sympy \footnote{\url{https://www.sympy.org/en/index.html}} syntax, these entries are processed using sympy  to ensure their validity. Matrix elements can be symbolic or numeric. 

Parameters are specified as comma separated tuples of the form <parametername> = <value>,...  Again the value can be symbolic.

\textbf{Setting network properties}

\begin{figure}
    \centering
    \includegraphics[width=1\linewidth]{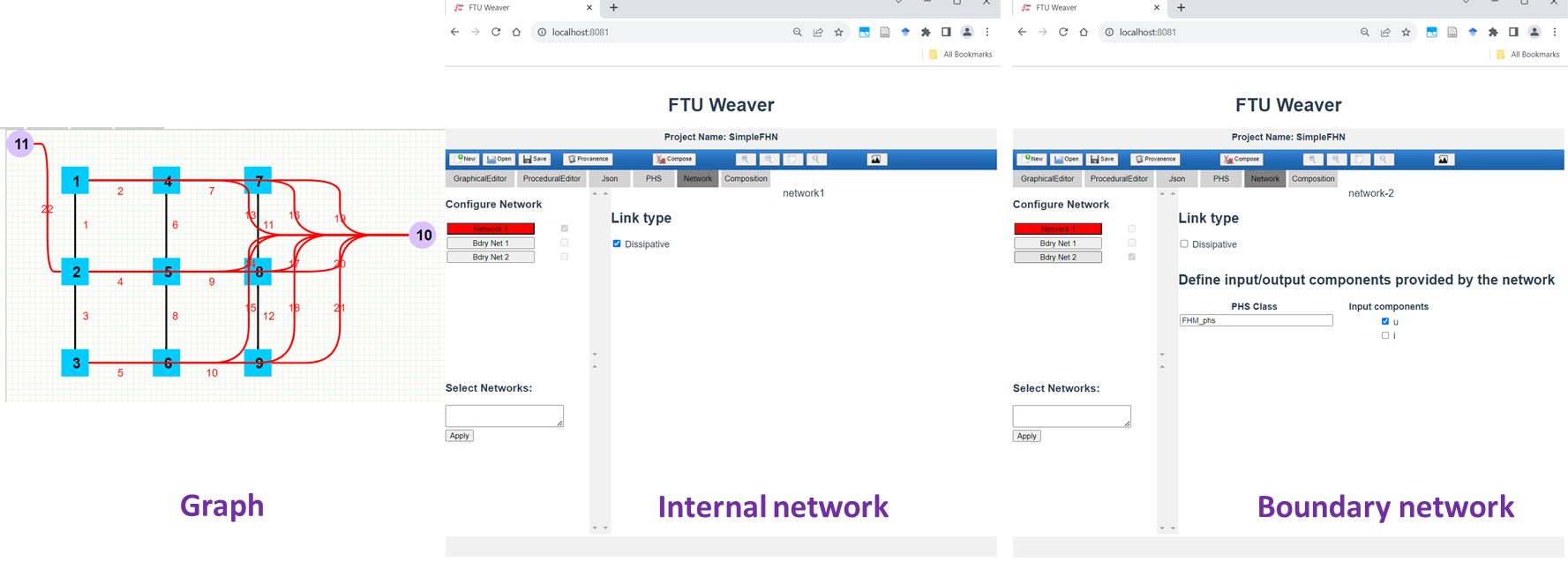}
    \caption{A simple network consisting of nine internal nodes, and two boundary nodes. The corresponding network properties in the \colorbox{gray!10}{Network} tab. All dissipative network entries have a red background. PHS class selection is available for boundary networks.}
    \label{fig:networkeditorui}
\end{figure}

The Network Editor tab, Figure \ref{fig:networkeditorui}, lists all the networks that have been defined on the graph. Note that, each boundary nodes defines a network. This interfaces enables the user to specify dissipative nature of the networks (i.e whether it is dissipative or non-dissipative). 

Boundary networks require input-vector component to which the associated boundary node couples with (phs class is not assigned to boundary nodes). This is specified by selecting the PHS entry from the dropdown list. Once a class/entry is selected, the input vector is listed. Users are expected to select the input-vector component from this list.

The \textbf{Select Networks} control allows users to select one or more networks based on their ids. Multiple networks can be specified by listing their comma separated ids. Note that boundary networks have negative ids.

\subsection{FTU composition specification workflow}

\begin{enumerate}
    \item Create the FTU graph, load/create and assign PHS entries to the nodes, specify edge weights (if there are dissipative networks), and complete the specification of network properties,
    \item Provide provanence information by clicking the \colorbox{gray!10}{Provanence} control icon on the control bar, this could done prior to creating the graph,
    \item Save the project by clicking the \colorbox{gray!10}{Save} control icon on the control bar, this could done during the graph creation process to save progress,but require provanence information to be apriori provided. The saved project in json format is available (readonly) in the \colorbox{gray!10}{JSON} Editor tab,
    \item Validate the inputs by composing the FTU specification by clicking the \colorbox{gray!10}{Compose} control icon on the control bar. This steps checks the input and creates a FTU composition. It will alert the user if the composition is incomplete or there are errors. The output of this process is  made available in the \colorbox{gray!10}{Composition} Editor tab. In case of no errors, a 'Success' message is displayed and a control is enabled to save the composition in json format.  This json is self contained and used in subsequent simulation and analysis workflows.
\end{enumerate}

\section{Next steps}
Tools and workflows to analyse the composite FTU and to extract useful information about the dynamics, perform model order reduction etc. are currently underdevelopment.
Given the variety of approaches that modellers may undertake to develop FTUs, flexibility to personalise the tools and workflows are necessary. We hope to meet these requirements as teams start using the interface and provide feedback on how the plan to use the composite FTUs.

\bibliographystyle{unsrtnat}
\bibliography{references} 

\end{document}